\documentclass[10pt]{IEEEtran}
\usepackage{graphicx}
\usepackage[cmex10]{amsmath}
\begin{document}

\title{Nonlinear Near-Field Microwave Microscope For RF Defect Localization in Superconductors}

\author{Tamin Tai, X. X. Xi, C. G. Zhuang, Dragos I. Mircea, Steven M. Anlage

\thanks{Manuscript received August 3, 2010. This work is supported by Department of Energy/High Energy Physics.}%
\thanks{Tamin Tai and Steven M. Anlage are with the Center for Nanophysics and Advance Materials,
Physics Department, University of Maryland, College Park, MD 20742
USA (email:tamin@umd.edu).}%
\thanks{X. X. Xi and C. G. Zhuang are with the Department of Physics, Temple University,
Philadelphia, PA 19122 USA}%
\thanks{Dragos I. Mircea is with Western Digital Media, Inc. 1710 Automation Parkway San Jose, CA,
95131 USA }}
 \maketitle

\begin{abstract}
Niobium-based Superconducting Radio Frequency (SRF) cavity
performance is sensitive to localized defects that give rise to
quenches at high accelerating gradients. In order to identify these
material defects on bulk Nb surfaces at their operating frequency
and temperature, it is important to develop a new kind of wide
bandwidth microwave microscopy with localized and strong RF magnetic
fields. By taking advantage of write head technology widely
used in the magnetic recording industry, one can obtain $\sim$200 $mT$ RF
magnetic fields, which is on the order of the thermodynamic critical field of Nb, on
sub-micron length scales on the surface of the superconductor. We have
successfully induced the nonlinear Meissner effect via this magnetic
write head probe on a variety of superconductors. This design should have a high spatial
resolution and is a promising candidate to find localized defects on
bulk Nb surfaces and thin film coatings of interest for accelerator
applications.
\end{abstract}


\section{Introduction}
\IEEEPARstart{S}{uperconducting} Radio Frequency (SRF) cavities will be used in the International Linear Collider (ILC) to explore
electron-positron collisions in high energy physics research. In
order to achieve a 1 TeV beam energy it is necessary to build $\sim 10^4$ Nb cavities with electrodynamic properties approaching the intrinsic limit dictated by theory. Despite the maturity of material fabrication techniques and improvement of
chemical and physical surface treatments and annealing processes in the
past several years, it is still challenging to fabricate so many state-of-the-art Nb cavities without performance-limiting defects. In general, many types of defects are
found on Nb cavity surfaces. Under intense RF loading, some of these defects can act as a hot spot to locally warm up the Nb superconductor
above its critical temperature ($T_{c}$), leading to a quench of the cavity.

One approach to this problem is to postpone
the quench of the superconductor by enhancing the RF breakdown field
of the material at the surface. There is considerable interest in preparing novel coatings on Nb cavities. Superconductor/insulator multilayer thin
film coatings have been proposed to enhance the RF breakdown field of the
superconductor~\cite{A. Gurevich}. It is of interest to measure whether or not this enhancement is possible with practical materials.

However, the properties of uncontrolled localized defects present in
the finished SRF cavities appear to limit their ultimate microwave
performance. Therefore, there is an urgent need to understand the
connections between localized defects, surface treatments, and the
RF breakdown field in the high frequency regime. Optical microscopy
techniques have been developed to identify defects in finished Nb
cavities~\cite{Y. Iwashita}. However this optical screening process
may result in identification of relatively benign defects which
will not result in a quench of the superconductor. Ideally, one
would like a microscopic technique that identifies defects based on
their poor microwave performance at low temperatures in the
superconducting state. One of the best candidates for this job is
the near field microwave microscope which has been developed
to quantitatively image RF and microwave properties of a variety of
materials on deep sub-wavelength scales~\cite{S. M.
Anlage},~\cite{D. I. Mircea}.

In order to generate a strong and localized RF magnetic field, and
to enhance the spatial resolution of this microscope, a magnetic
writer is utilized in our experiment. Taking advantage of magnetic
write head technology with write-gap widths on the order of 100
nm~\cite{K. Z. Gao}, an RF field on the scale of 1 Tesla~\cite{Wang}
with sub-micron spatial extent~\cite{Kob} can be created. In this
work, we integrate the magnetic writer probe into our microwave
microscope and demonstrate that this probe can develop a nonlinear
Meissner effect signal from several kinds of superconductors such as
MgB$_{2}$ and Tl$_{2}$Ba$_{2}$CaCu$_{2}$O$_8$ (TBCCO). This probe
has great potential for high resolution nonlinear Meissner effect
microscopy and in the future will be used in analyzing defects on Nb cavity
surfaces at high frequencies and low temperatures.

\section{EXPERIMENT}
\begin{figure}[!t]
\centering
\includegraphics[height=1.4 in, width=2.7in]{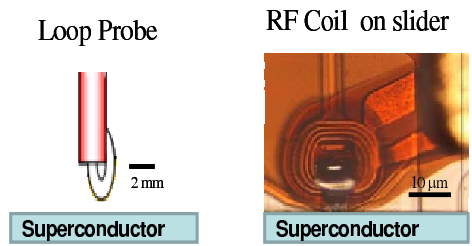}
\caption{Comparison of loop probe (left) and magnetic write
head probe (right).  A 4-turn coil is visible inside this magnetic
write head, which develops a high frequency magnetic field in a write-gap near the surface of the superconductor.} \label{fig_probe}
\end{figure}

\subsection{Experimental Setup}
In previous work, we developed a low resolution near field microwave
microscope to nondestructively measure the local harmonic generation
from unpatterned superconducting samples ~\cite{D. I.
Mircea},~\cite{S. C. Lee},~\cite{Lee S. C.},~\cite{Lee2}. In this
design, a loop probe (Fig.~\ref{fig_probe} left) is made by shorting
the inner conductor and outer conductor of a commercial semi-ridge
coax cable with inner diameter 200 $\mu$$m$ and outer diameter 2
$mm$.  The loop is brought to within 10 $\mu$$m$ of the
superconducting surface and a high frequency signal is applied to
the loop.  RF screening currents are induced in the sample.  Due to
its nonlinear response, harmonics of the drive signal are created,
and these couple back into the loop probe and are measured at room
temperature with a spectrum analyzer. This experiment can determine
both the second and third harmonic nonlinear products at the same
location at any excited frequency and temperature of interest.

Here, we have modified this basic experiment to produce stronger and more
localized RF magnetic fields. Based on the need to investigate Nb
near the thermodynamic critical field, we require at least 200
${mT}$ magnetic field at the sample surface. In order to enhance the
magnetic coupling between probe and superconducting sample, we
replace the loop probe with a magnetic write head
(Fig.~\ref{fig_probe} right).

\begin{figure}[!t]
\centering
\includegraphics[height=3.5in, width=3.2in]{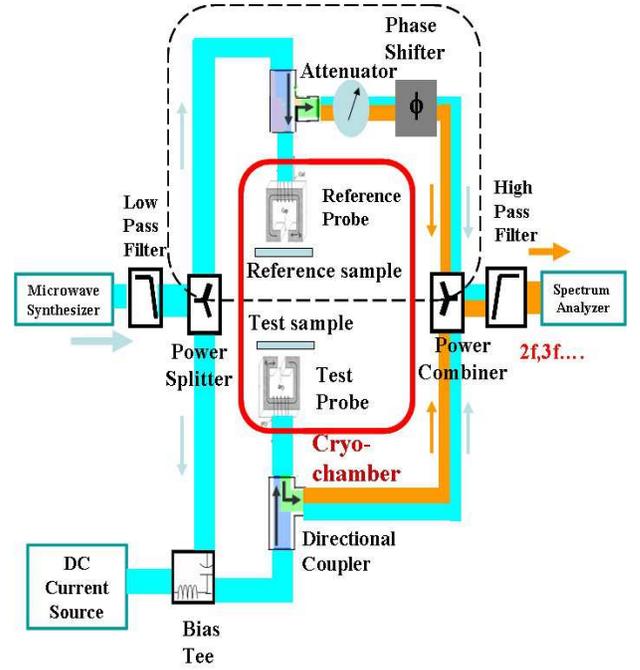}
\caption{Experimental setup. An excited wave (fundamental signal)
at approximately 3.5 GHz is low-pass filtered to eliminate
higher harmonics and sent to the tip of the microwave test probe. The
tip can be either a loop probe or magnetic write head probe. The
dashed line encloses a reference arm setup only used for
background signal cancellation.
A DC current can be injected into the microwave circuit via the
bias tee. Higher order harmonic signals
induced in the superconductor are gathered by the probe tip and
high pass filtered before being measured by the spectrum analyzer.}
\label{fig_setup}
\end{figure}

The system setup is shown in Fig.~\ref{fig_setup}. Both the magnetic
write head probe and the superconductor are kept in a high vacuum
cryogenic environment. Microwave fundamental frequency power $P_{f}$
is generated by a microwave synthesizer. Low pass filters are used
to filter out higher harmonics generated by the microwave source. Because of the
perturbation of the super fluid density coming from the externally
applied RF magnetic field, higher order harmonic response ($P_{2f}$,
$P_{3f}$, ${\cdot\cdot\cdot}$) will be induced on the sample for
temperatures below $T_{c}$. Those harmonic signals will
be radiated from the sample and can be extracted by high pass
filtering the signal from the probe. Here we shall concentrate on
$P_{3f}$, which arises from time-reversal invariant perturbations of
the superconductor and can be used to examine both intrinsic and
extrinsic nonlinear characteristics of the material.

In Fig.~\ref{fig_setup}, the closed dashed line encloses a reference microwave circuit designed for cancellation of nonlinearity from the magnetic
write head probe itself, and will be discussed below. In addition, a
bias tee is integrated into the microwave microscope circuit to
allow injection of a small DC current into the probe, also discussed further below.

The generated third harmonic power $P_{3f}$ is estimated as~\cite{D.
I. Mircea},~\cite{Lee2}

\begin{equation}\label{P3f:equation}
    P_{3f} \propto \frac{\omega^2\lambda^4(T)\Gamma^2}{J^4_{NL}(T,x)}
    \qquad
    \Gamma=\int\frac{K^4(x,y)dxdy}{I_{total}}
\end{equation}
 where $\omega$ is the frequency of the incident wave, $\lambda (T)$ is the temperature dependent magnetic
penetration depth, $J_{NL}$ and $\Gamma$ are the nonlinear scaling
current and a current-distribution geometry factor, respectively.
$K(x,y)$ is the surface current induced in the superconductor at the fundamental frequency and
${I_{total}}$ is total volume current induced in the sample. From
Eq.(~\ref{P3f:equation}), a strong magnetic field from the magnetic
write head probe will enhance the surface current $K$ and confine
the current distribution, both leading to an enhancement of
$\Gamma$, and therefore $P_{3f}$, for a given excitation power. This
has the added benefit of improving the spatial resolution of the
probe. Defective regions of the sample, including those responsible
for hot-spot generation, have smaller values of $J_{NL}$ than the surrounding material, hence will
develop larger $P_{3f}$, thus giving away their position to the
microscope. Previous work has demonstrated the ability of this
microscope to identify a grain boundary Josephson junction defect in
a cuprate thin film~\cite{S. C. Lee},~\cite{Lee S. C.},~\cite{Lee2}.

\subsection{Sample}
The superconducting samples we study include a TBCCO thin film
of thickness 500 $nm$,
epitaxially grown by the magnetron sputtering method~\cite{D. W. Face}.
In addition, a high quality epitaxial MgB$_{2}$ thin film with thickness
25 $nm$ is also examined.  The MgB$_{2}$ film is
deposited on a SiC substrate by a hybrid physical-chemical vapor
deposition technique~\cite{X. Zeng}. Both samples are examined at a single location in the center of the 10 mm*10 mm
area.

\section{Data and Discussion}

\subsection{ Magnetic Write Head Behavior at Microwave Frequencies}

\begin{figure}[!t]
\centering
\includegraphics[height=3in,width=3in]{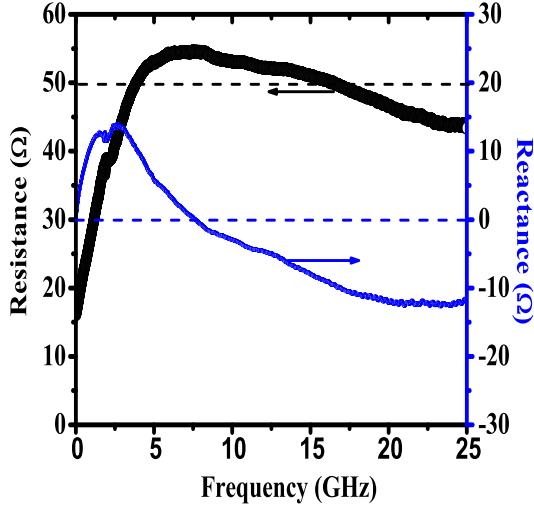}
\caption{Measured impedance of the magnetic write head as a function of frequency.  The thick line indicates the
resistance, and the thin line indicates the reactance values. The measurement is done with a PNA-X N5242A network
analyzer and Picoprobe at room temperature.} \label{fig_Impeadance}
\end{figure}

Before integrating the magnetic write head into our near field
microwave microscope, we measured the complex load impedance that the head presents to the microwave generator.  Fig.~\ref{fig_Impeadance} shows the impedance measured with a Picoprobe touching the contact pads on the slider. Remarkably, the write head is very well impedance matched to 50 $\Omega$ in
resistance and 0 $\Omega$ in reactance over a broad frequency range
from around 2 GHz to 25 GHz, which is quite ideal for the present application. Such good impedance match implies that we can deliver 45 $mA$ of current to the write head using 100 $mW$ (+20 dBm) of RF power, and we have found that this does not burn out the magnetic write head coil.

\subsection{Third Order Nonlinear Response from Superconducting Samples}

\begin{figure}[!t]
\centering
\includegraphics[height=3in,width=3.4in]{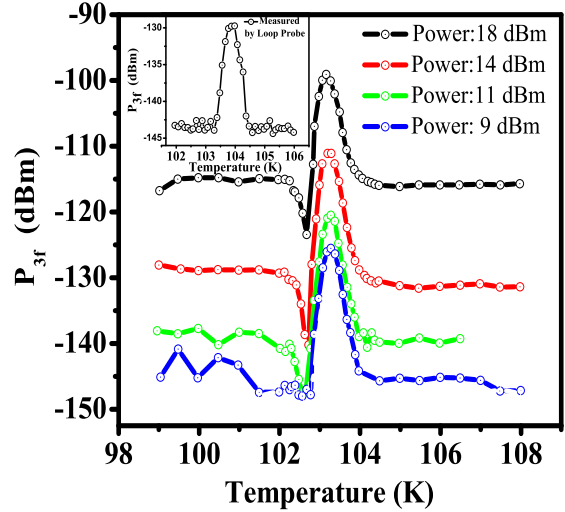}
\caption{Temperature dependence of third harmonic power P$_{3f}$ of a TBCCO film, measured
by a magnetic write head probe with an excitation frequency of 3.5 GHz.
Inset is measured by the bare loop probe with excitation frequency of
3.5 GHz and excited power 18 dBm.} \label{fig_TBCCO}
\end{figure}

A measurement of the temperature dependent 3rd order harmonic power is
performed at the center of the TBCCO film by the magnetic write
head probe with different exciting powers. The inset of
Fig.~\ref{fig_TBCCO} shows $P_{3f}(T)$ measured by the bare
loop probe.  A peak in $P_{3f}$ near $T_{c}$ shows up, as expected.
This enhancement of $P_{3f}$ is due to the nonlinear Meissner
effect near $T_{c}$.  From Eq.
\ref{P3f:equation} one sees that due to $J_{NL}$ approaching zero and
$\lambda(T)$ diverging at $T_{c}$ the third harmonic power will increase strongly.  The divergence is cut off by the distribution of transition temperatures in the sample, and the influence of quasiparticle electrodynamics. With the loop probe,
the enhancement of $P_{3f}$ above background is only 15 dB, for 18 dBm fundamental input power. Such a small enhancement can be easily achieved by the magnetic head probe with only 9 dBm excited power (Fig.
~\ref{fig_TBCCO}), which means that the magnetic write head generates a more localized and intense field, inducing stronger surface currents on the sample.  Despite these higher currents, there is no evidence of localized heating in the sample from the data in Fig.~\ref{fig_TBCCO}.

\begin{figure}[!t]
\centering
\includegraphics[height=3in, width=3 in]{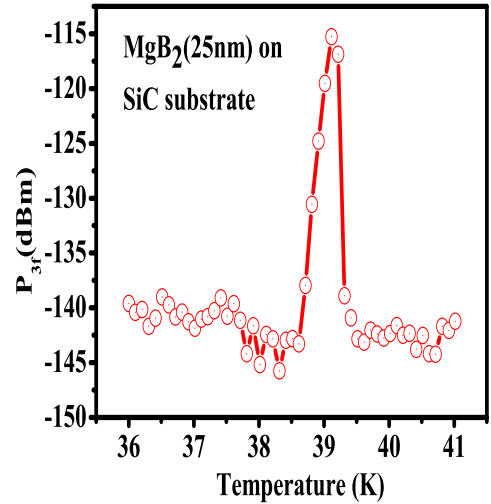}
\caption{Temperature dependence of third harmonic power P$_{3f}$ from an MgB$_{2}$ film,
measured with the magnetic write head probe with an excitation  frequency
of approximately 3.5 GHz and a power of 12 dBm.} \label{fig_MgB2}
\end{figure}

In order to test the magnetic write
head probe in a liquid Helium cooled
environment, temperature dependent 3rd order harmonic power is also
measured in the center position of an MgB$_{2}$ thin film. In Fig.
~\ref{fig_MgB2}, a peak at 39.1 K shows up clearly near the $T_{c}$
of the film. This proves that the magnetic probe can function
in the low temperature region. Comparison of this peak
with that of TBCCO, one finds a much sharper transition, implying a narrow distribution of $T_c$ values in the MgB$_{2}$ thin film.

\subsection{Cancellation of Nonlinearity from the Probe}

\begin{figure}[!t]
\centering
\includegraphics[height=2.8in,width=3.4 in]{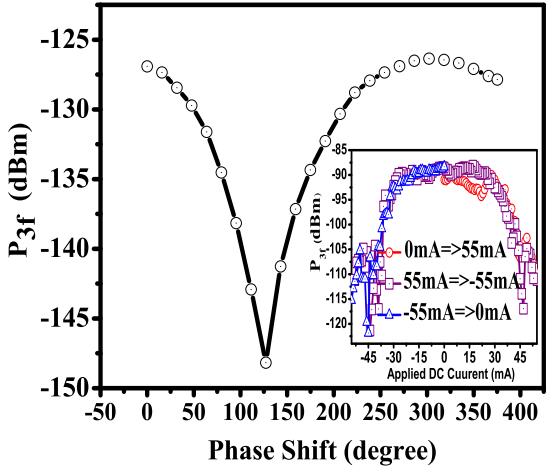}
\caption{Measurement of total third harmonic power $P_{3f}$ from the sample probe and reference probe, as a function of phase shift in the full setup shown in Fig.~\ref{fig_setup}.  The experiment is carried out with an
excitation frequency of 2.12 GHz and power of 14 dBm at ambient
temperature under vacuum.   No samples are present in this measurement.}
\label{fig_phase_cancelation}
\end{figure}

One can see from Fig.
~\ref{fig_TBCCO} that as the excitation power to the probe is increased, the noise floor is also enhanced. This is due to the nonlinearity of the
magnetic write head itself. Generally speaking, all of the tested magnetic write heads show some degree of nonlinearity. To clarify the
origin of this probe nonlinearity, a DC current is
injected into the write head drive coil to control the magnetization
direction of the magnetic materials inside the probe.  Third harmonic power from the probe under 3.5
GHz and 16 dBm illumination is shown in the inset of Fig.~\ref{fig_phase_cancelation}, as a function of DC current from -55 mA to +55 mA.  The $P_{3f}$ from the magnetic write head decreases dramatically at +45 mA and -45 mA, demonstrating that an applied DC current can suppress background nonlinearity.  This decrease in $P_{3f}$ may be due to establishment of a fully magnetized state of the ferromagnetic films in the write head, thus elimianting nonlinearity from minor hysterisis loops.  The applied current may also reduce magnetic domain wall motion and magnetic moment precession, thus reducing background
nonlinearity~\cite{S. Y. An}.

To experimentally reduce the background nonlinearity, a reference
arm is created to cancel the contributions of the probe to the
measured nonlinearity.  The area circled by the dashed line in
Fig.~\ref{fig_setup} includes an identical magnetic write head
probe.  The phase shifter and variable attenuator are used to create
an equal amplitude but 180-degree phase-shifted third harmonic signal
from the reference arm.  A plot of the total $P_{3f}$ generated by
both arms combined, at different phase shifts, is shown in
Fig.~\ref{fig_phase_cancelation}. With 127.2$^{o}$ of phase shift,
the background nonlinearity is completely cancelled (down to the noise level of the spectrum analyzer) by the reference probe.  This result implies that the
microscope can be made sensitive to just the $P_{3f}$ signal from
the superconductor, despite the presence of nonlinear magnetic
materials in the write probe.  With this modification we expect to
achieve high spatial resolution in nonlinear near field microwave
microscopy, and apply it to defect identification in Nb materials.

At present the cryostat housing this microscope is not able to achieve sample temperatures below the transition temperature of Nb.  The probe assembly causes localized heating on the Nb sample, preventing it from being cooled below $T_{c}$.  Further improvements in the probe cooling system are required.

\section{Conclusion}
A magnetic write head is successfully integrated into the near field
microwave microscope operating at cryogenic temperatures. The magnetic write head should generate RF magnetic field on the scale of the thermodynamic critical field of Nb on sub-$\mu$m length scales. Using this probe, a clear reproducible nonlinear response signal from superconducting samples of TBCCO and MgB$_{2}$ are obtained. Although this probe can generate strong nonlinearity itself, a phase
cancellation method is demonstrated to nearly zero out this contribution. This microscope will be employed to identify defects that degrade the RF performance of Nb used in SRF cavities.


\begin{thebibliography}{1}

\bibitem{A. Gurevich}
A. Gurevich, ``Enhancement of RF breakdown field of superconductors
 by multilayer coating," \emph{Appl. Phys. Lett.}, vol. 88, pp. 012511, 2006.

\bibitem{Y. Iwashita}
Y. Iwashita, Y. Tajiro, H. Hayano, ``Development of high resolution
camera for observations of superconducting cavities," \emph{Phys.
Rev. ST Accel. Beams}, vol. 11, pp. 093501, 2008.

\bibitem{S. M. Anlage}
S. M. Anlage, V. V. Talanov, A. R. Schwartz, \emph{Principles of
Near-Field Microwave Microscopy in Scanning Probe Microscopy:
Electrical and Electromechanical Phenomena at the Nanoscale}, New
York: Springer-Verlag, vol. 1, pp. 215-253, 2007.

\bibitem{D. I. Mircea}
D. I. Mircea, H. Xu, S. M. Anlage, ``Phase-sensitive harmonic
measurements of Microwave Nonlinearities in Cuprate Thin Films,"
\emph{Phys. Rev. B}, vol. 80, pp. 144505, 2009.

\bibitem{K. Z. Gao}
K. Z. Gao, O. Heinonen, Y. Chen, ``Read and write processes, and
head technology for perpendicular recording", \emph{J. Magn. Magn.
Mater.}, vol. 321, pp. 495-507, 2009.

\bibitem{Wang}
 S. X. Wang and A. M. Taratorin, \emph {Magnetic Information Storage
Technology}, Academic Press, San Diego, 1999, p. 89.

\bibitem{Kob}
M. R. Koblischka, J. D. Wei, M. Kirsch, U. Hartmann, ``High
frequency magnetic force microscopy-imaging of harddisk write heads",
 \emph{Japanese J. Appl. Phys.} vol. 45, pp. 2238-2241, 2006.

\bibitem{S. C. Lee}
S.-C. Lee, S. M. Anlage, ``Study of Local Nonlinear Properties Using
a Near-Field Microwave Microscope," \emph{IEEE Trans. Appl.
Supercond.}, vol. 13, pp. 3594-3597 2003.

\bibitem{Lee S. C.}
S.-C. Lee, S. M. Anlage, ``Spatially resolved nonlinearity
measurements of YBa$_{2}$Cu$_{3}$O$_{7}$  bi-crystal grain
boundaries," \emph{Appl. Phys. Lett.}, vol. 82, pp. 1893-1895, 2003.

\bibitem{Lee2}
S.-C. Lee, S.-Y. Lee, S. M. Anlage, ``Microwave Nonlinearities of an
Isolated Long YBa$_{2}$Cu$_{3}$O$_{7-\delta}$  Bicrystal Grain
Boundary," \emph{Phys. Rev. B}, vol. 72, pp. 024527, 2005.

\bibitem {D. W. Face}
D. W. Face, R. J. Small, M. S. Warrington, F. M. Pellicone, P. J.
Martin. ``Large area YBa$_{2}$Cu$_{3}$O$_{7}$ and
Tl$_{2}$Ba$_{2}$CaCu$_{2}$O$_{8}$ thin films for microwave and
electronic applications." \emph{Physica C}, vol. 357, pp. 1488-1494,
2001.

\bibitem {X. Zeng}
X. Zeng, A. V. Pogrebnyakov, A. Kotcharov, J. E. Jones, X. X. Xi, E.
M. Lysczek, J. M. Redwing, S. Y. Xu, J. Lettieri, D. G. Schlom, W.
Tian, X. Q. Pan, Z. K. Liu, ``In situ epitaxial MgB$_{2}$ thin films
for superconducting electronics," \emph{Nature Materials}, vol. 1,
pp. 35-38, 2002.

\bibitem {S. Y. An}
S. Y. An, P. Krivosik, M. A. Kraemer, H. M. Olson, A. V. Nazarov, C.
E. Patton, ``High power ferromagnetic resonance and spin wave
instability processes in Permalloy thin films", \emph{J. Appl.
Phys.}, vol. 96, pp.1572-1580, 2004.

\end{thebibliography}
\end{document}